\documentstyle[12pt]{article}

\renewcommand{\thefootnote}{\fnsymbol{footnote}}

\renewcommand{\baselinestretch}{1.24} 
\makeatletter

\@addtoreset{equation}{section}
\makeatother

\begin{document}

\begin{titlepage}
\hfill\parbox{4cm}
{hep-th/0211217}

\vspace{20mm}
\centerline{\Large \bf Chiral primary cubic interactions} 
\vspace{3mm}
\centerline{\Large \bf from pp-wave supergravity}

\vspace{10mm}
\begin{center}
Youngjai Kiem$^a$\footnote{ykiem@muon.kaist.ac.kr},
Yoonbai Kim$^{b,c}$ \footnote{yoonbai@skku.ac.kr},
Jaemo Park$^d$\footnote{jaemo@physics.postech.ac.kr},
and Cheol Ryou$^a$\footnote{cheol@muon.kaist.ac.kr}
\\[10mm]
{\sl $^a$ Department of Physics, KAIST, Taejon 305-701, Korea}
\\[2mm]
{\sl $^b$ BK21 Physics Research Division and Institute of Basic
Science}
\\
{\sl Sungkyunkwan University, Suwon 440-746, Korea}
\\[2mm]
{\sl $^c$ School of Physics, Korea Institute for Advanced Study,}
\\
{\sl 207-43, Cheongryangri-Dong, Dongdaemun-Gu, Seoul 130-012, Korea}
\\[2mm]
{\sl $^d$ Department of Physics, POSTECH, Pohang 790-784, Korea}

\end{center}

\thispagestyle{empty}
\vskip 22mm

\begin{abstract}
\noindent
We explicitly construct cubic interaction light-cone 
Hamiltonian for the chiral primary system involving the 
metric fields and the self-dual four-form fields in the 
IIB pp-wave supergravity.  The background fields 
representing pp-waves exhibit
SO(4)${}_{\perp} \times$SO(4)${}_{\parallel} \times Z_2$ 
invariance.  It turns out that the interaction Hamiltonian 
is precisely the same as that for the dilaton-axion system, 
except for the fact that the chiral primary system fields 
have the opposite parity to that of the dilaton-axion 
fields under the $Z_2$ transformation that exchanges 
two SO(4)'s.
\end{abstract}

\vspace{2cm}
\end{titlepage}

\baselineskip 7mm
\renewcommand{\thefootnote}{\arabic{footnote}}
\setcounter{footnote}{0}


While the IIB string theory in flat space-time
exhibits SO(8) invariance, pp-wave backgrounds of
Refs.~\cite{bbb,bmn} possess 
SO(4)${}_{\perp} 
\times$SO(4)${}_{\parallel} \times Z_2$
invariance.  The consequence of this fact at the
level of supergravity spectrum is that there
are two extra scalars $s$ and $\bar{s}$
under SO(4)${}_{\perp} \times$ SO(4)${}_{\parallel}$
on top of the dilaton and axion that are
scalars under SO(8) ($\tau$
and $\bar{\tau}$ with the
energy $m^2 = 4 f$).  They come from
the trace part of the SO(4)${}_{\perp}$ subspace
metric, that is related to the SO(4)${}_{\parallel}$
subspace metric via the SO(8) traceless
condition, and the four-form RR gauge
field along the SO(4)${}_{\perp}$, which is related
to the SO(4)${}_{\parallel}$ part through the self-duality
condition.  At the free theory level, it is known
that they combine to produce chiral primary with
$m^2 = 0$ and anti-chiral primary with
$m^2 = 8 f$ \cite{metstey}.  Recalling that the $Z_2$ part of the 
symmetry is an element of the SO(8) that
exchanges SO(4)${}_{\perp}$ and 
SO(4)${}_{\parallel}$,
the dilaton and axion, being SO(8) scalars, have
even parity under the $Z_2$.  When viewed from this 
angle, chiral primaries should have odd parity.  The
confirmation of this expectation at the supergravity
interaction level is the purpose of this paper. 
Specifically, we construct the cubic interaction part of the
light-cone Hamiltonian for the chiral primary
system starting from the covariant IIB supergravity.  
It turns out that thus obtained
interaction Hamiltonian is precisely the same
as that for the dilaton-axion system \cite{sug,sug1}, except
for the fact that 
$s$ and $\bar{s}$ ($\tau$ and $\bar{\tau}$)
have odd (even) parity under the
exchange of the two SO(4)$'$s\footnote{See \cite{parity} for
related discussions.}.  This result
is interesting at least from two perspectives.
First, in view of the apparent difference
in the SO(8) tensor structure of the two systems,
this agreement is not entirely trivial. 
Generically, the treatment of the self-dual
four-form fields in curved backgrounds is
a non-trivial problem \cite{ss}, but the pp-wave backgrounds
provide us with a tractable setting.  Secondly,
the ground state\footnote{In the string field theory 
side \cite{str,str1},
this state is the supersymmetric ground state $|vac\rangle$
satisfying $Q|vac\rangle = \bar{Q}|vac\rangle = 0$ with
zero energy.  
In the dual gauge theory side \cite{ym,mix}, this corresponds to the
operator ${\rm Tr} Z^J$ of Ref.~\cite{bmn}.  The 
`string states' in the weak coupling limit of 
the dual gauge theory has the energy 
(slightly) larger than $2f$.} 
of the string theory in pp-wave backgrounds should be given
odd parity; supergravity can be obtained from
string theory via a smooth deformation of 
parameters, and the discrete charge assignment
should not change under this deformation.

%

The bosonic sector of type IIB supergravity
involves a dilaton, an axion, a graviton, two antisymmetric 
2-form fields, and a self-dual antisymmetric four-form field.
We are interested in the system of the gravity field 
$g_{\mu\nu}$  and the self-dual antisymmetric 4-form gauge field
$a_{\mu\nu\rho\sigma}$.  The classical dynamics of these fields 
are described by Einstein equations  
\begin{equation}\label{eineq}
R_{\mu\nu}=\frac{1}{4!}F_{\mu\rho\sigma\tau\lambda}
F^{\;\;\rho\sigma\tau\lambda}_{\nu} ~ ,
\end{equation}
and the four-form gauge field equations
\begin{equation}\label{maxeq}
\nabla_{\mu}F^{\mu\nu\rho\sigma\tau}=0 ~ ,
\end{equation}
where the five-form field strength tensor is defined as
\begin{equation}\label{5fo}
F_{\mu\nu\rho\sigma\tau}\equiv 
\partial_{[\mu}a_{\nu\rho\sigma\tau]} ~ ,
\qquad \mu,\nu,...=+,-,1,2,\cdots ,8 ~ .
\end{equation}
In addition, the self-duality condition \cite{metstey}
\begin{equation}\label{sdc}
F_{\mu\nu\rho\sigma\tau}=-\frac{1}{5!}\sqrt{-g}
\epsilon_{\mu\nu\rho\sigma\tau\mu'\nu'\rho'\sigma'\tau'}
F^{\mu'\nu'\rho'\sigma'\tau'} ~ ,
\end{equation}
where 
\begin{equation}
\qquad \epsilon_{+-12345678}=-1 , ~~~ 
 \epsilon_{0123456789}= - 1 ~ ,
\end{equation}
should be imposed.  In fact, it is well known that the
self-duality condition (\ref{sdc}) implies the equations
of motion (\ref{maxeq}).


To obtain the pp-wave backgrounds, we turn on constant 
background  condensate $f$ for certain components of the
five-form field strength 
\begin{equation}\label{cond}
F_{+1234}=F_{+5678}=2f ~,
\end{equation}
and set others to zero.  The nontrivial component of the 
Einstein equations (\ref{eineq}) is
\begin{equation}
R_{++}=8f^{2} ~,
\end{equation}
solving which determines the metric to be
\begin{equation}\label{ppmet}
ds^{2}=2dx^{+}dx^{-}-f^{2}x^{2}_{I}dx^{+}dx^{+}+dx^{I}dx^{I} ~ ,
\end{equation}
where $I,J,...=1,2,\cdots ,8$ denote SO(8) indices. The pp-wave 
metric (\ref{ppmet}) and the constant condensate for the field 
strength (\ref{cond}) constitute our background configuration 
in what follows.


Our analysis of the gravity sector will closely follow that
of Goroff and Schwarz~\cite{gor}.  Adopting their notations,
the metric fluctuations are parametrized as follows: 
\begin{equation}\label{mmet}
(g)_{\mu\nu}=
\left(
\begin{array}{ccc}
g_{++} & g_{+-} & g_{+J} \\
g_{-+} & g_{--} & g_{-J} \\
g_{I+} & g_{I-} & g_{IJ}
\end{array}
\right),
\end{equation}
where 
\begin{equation}\label{grga}
g_{++}=-f^{2}x^{2}_{I}+h_{++}, \qquad g_{+-}=e^{\varphi}, 
\qquad g_{IJ}=e^{\psi}\gamma_{IJ}~\mbox{with}~ 
\det(\gamma_{IJ})=1 ~.
\end{equation}
To implement the light-cone gauge fixing, we choose the
nine components of the metric to zero
\begin{equation}\label{ggc}
g_{--}=g_{-I}=0 ~ .
\end{equation}
Using the remaining gauge invariance out of the ten 
original
diffeomorphisms, we impose a relation 
between $g_{+-}$ and $\det(g_{IJ})$
\begin{equation}\label{ppi}
 \varphi=\frac{1}{2}\psi ~. 
\end{equation}
This condition simplifies the $(--)$-component of Ricci 
tensor
\begin{equation}\label{rmm}
R_{--}=4\left[\partial^{2}_{-}\psi+ \frac{1}{2} (\partial_{-}\psi)^{2}
-\partial_{-}\varphi\partial_{-}\psi\right]-\frac{1}{4}
\partial_{-}\gamma_{IJ}\partial_{-}\gamma^{IJ} ~ ,
\end{equation}
where $\gamma^{IJ}$ satisfies 
$\gamma^{IJ}\gamma_{JK}=\delta^{I}_{\;K}$.
Since we are interested in the chiral primary system,
we turn on only one gravitational fluctuation out of the
35 physical graviton degrees of freedom 
$\gamma_{IJ}$
\begin{equation}\label{gaij}
\gamma_{IJ}=\left(
\begin{array}{cc}
e^{h}I_{4} & 0 \\
0 & e^{-h}I_{4}
\end{array}
\right),
\end{equation}
where $I_{4}$ is $4\times 4$ unit matrix.  This is the
general form of the physical field fluctuations satisfying
$\det(\gamma_{IJ})=1$ that are
scalars under SO(4)${}_\perp \times$SO(4)${}_\parallel$.
Note that the field $h$ has negative parity under the
$Z_2$ transformation that exchanges the two SO(4)'s, namely,
under the interchange of the upper and lower blocks.
The expression for the inverse metric
\begin{equation}\label{mnup}
(g)^{\mu\nu}=
\left(
\begin{array}{ccc}
0 & e^{-\psi/2} & 0 \\
e^{-\psi/2} & (f^{2}x^{2}_{I}-h_{++})e^{-\psi}
+g_{+I}\gamma^{IJ}g_{J+} & -e^{-\frac{3}{2}\psi}
g_{+I}\gamma^{IJ} \\
0 & -e^{-\frac{3}{2}\psi} g_{+J}\gamma^{JI} & e^{-\psi}\gamma^{IJ}
\end{array}
\right) 
\end{equation}
will be useful in what follows.

For the self-dual four-form gauge field $a_{\mu\nu\rho\sigma}$, 
we choose the light-cone gauge condition
\begin{equation}\label{lcg}
a_{-\mu\nu\rho}=0 ~ ,
\end{equation}
which is possible due to the presence of the gauge
transformations for the four-form gauge field.
After the imposition of the self-duality condition (\ref{sdc}),
there are 35 physical degrees of freedom.  Among these,
we again turn on fluctuations
that are scalars under
 SO(4)${}_{\perp}\times$SO(4)${}_{\parallel}$ transformations. 
Consequently, all mixed components such as 
$a_{ijkl'}$ ($a_{ij'k'l'}$) or $a_{ijk'l'}$ vanish, where
$i, j , \cdots = 1, 2, 3, 4$ are vector indices of 
SO(4)${}_{\parallel}$ and $i' , j' , \cdots = 5, 6, 7, 8$ 
are vector indices of SO(4)${}_{\perp}$.  The nonvanishing
components $a_{1234}$ and $a_{5678}$ are
constrained by the  $(-1234)$-component (or, 
equivalently, $(-5678)$-component) of the self-dual 
equation (\ref{sdc})~:
\begin{equation}\label{sdc1}
\partial_{-}a_{1234}=-e^{4h}\partial_{-}a_{5678}. 
\end{equation}
In the derivation of (\ref{sdc1}), we have 
used $\sqrt{-g}=e^{9\psi/2}$ and the explicit form of
the upper index metric (\ref{mnup}).  We note that when
the relative sign between the $(+1234)$-component 
and the $(+5678)$-component in (\ref{cond}) is positive, 
the relative sign appearing
in (\ref{sdc1}) is necessarily negative.  Consequently, when 
we turn off the metric fluctuation $h$, the field $
\partial_- a_{1234}
= - \partial_- a_{5678} = \partial_- a $
has negative parity under the $Z_2$ transformation. 
 

Our ultimate goal is to identify the quadratic action for the
chiral primary system $(h,a)$ and to construct 
cubic interactions  among them.  We thus solve the gauge 
constraints from both the
Einstein equations (\ref{eineq}) and the self-dual equations (\ref{sdc})
to express the unphysical fields in terms of the physical
fields.  To be specific, 
$(--)$-, $(+-)$-, $(-I)$-components of the Einstein equations are 
solved to express ten auxiliary fields $(h_{++},g_{+I},\psi)$ of the
covariant metric (\ref{mmet}), or equivalently $(g^{--},g^{-I},\psi)$ 
of the contravariant metric (\ref{mnup}) in terms of $(h,a)$.
With the help of Eqs.~(\ref{ppi})--(\ref{gaij}), 
the Einstein equation (\ref{ppi}) for the variation of $g^{--}$ leads to
\begin{eqnarray}
R_{--}&=&4\partial^{2}_{-}\psi+2(\partial_{-}h)^{2}\nonumber\\
 & = & \left[(\partial_{-}a_{1234})^{2}+(\partial_{-}
 a_{5678})^{2} + {\cal O} (3) + \cdots \right] ~ ,
\end{eqnarray}
up to quadratic terms.  We thus obtain 
\begin{equation}\label{psii}
\psi= \frac{1}{4}\left(\frac{1}{\partial^{2}_{-}}\right)
\left[-2(\partial_{-}h)^{2}+(\partial_{-}a_{1234})^{2}
+(\partial_{-}a_{5678})^{2} \right]  + {\cal O} (3) + \cdots ~ . 
\end{equation}
The operator $1/ \partial_-$ is to be understood as the
inverse momentum in the momentum basis, as is conventional
in the light-cone quantization \cite{gor}.  We also mention
that one should carefully use integration by parts; for the 
cubic terms, for example, it is easier to understand it as 
using the momentum conservation conditions, such as
$p_1^+ = - ( p_2^+ + p_3 ^+ )$. 
There can in principle be subtle zero mode effects, which often play an 
important role in light-cone approaches 
(see, for example, \cite{yama}). 
Our final answer, however,
is Lorentz invariant when we set $f = 0$.  Noting the
fact that the terms involving $f$ also have 
the $\partial_-$ derivative and thereby undetectable by the
the zero mode parts, the demonstration of the Lorentz
invariance of the answer when $f=0$ should be sufficient.   

Up to quadratic terms, $(-I)$- and $(-+)$-components of the 
Einstein equations 
(\ref{eineq}) are 
\begin{eqnarray}
R_{-I}&= &\frac{1}{2}\partial_{-}^{2} 
\left(g^{-I}-\Delta_{IJ}\frac{\partial_{J}}{\partial_{-}}h\right)
+2 \partial_{-}h\partial_{I}h+\frac{15}{4}
\partial_{I}\partial_{-}\psi + {\cal O} (3) + \cdots
\nonumber\\
&= & \frac{1}{4!}\partial_{-}a_{JKLM}\partial_{I}a_{JKLM}
-\frac{1}{3!}\partial_{-}a_{+JKL}\partial_{-}a_{IJKL}  
 + {\cal O} (3) + \cdots ~ ,
\label{rmi}
\end{eqnarray}
where 
\begin{equation}
\Delta_{IJ}=\left(
\begin{array}{cc}
I_{4} & 0 \\
0 & -I_{4} 
\end{array}
\right)_{IJ} ~ .  \label{xxx}
\end{equation}
For brevity, we simply list the following expression for 
the metric $g^{-I}$ up to the linear terms
\begin{equation}\label{gj-}
g^{j-} = \frac{\partial_{j}}{\partial_{-}}h + {\cal O} (2) 
  + \cdots, \qquad
g^{j'-} = -\frac{\partial_{j'}}{\partial_{-}}h + 
{\cal O} (2) + \cdots ~ .
\end{equation}
In the case of $(+-)$-component, we solve
\begin{eqnarray}
R_{+-}& = & \frac{1}{2}\partial_{-}^{2}\left(g^{--}
-f^{2}x_{I}^{2}+\frac{\partial_{I}}{\partial_{-}}g^{-I}\right)
-\frac{1}{2}\left[\frac{1}{2}\partial_{-}^{2}(g^{-I}g^{-I}-2f^{2}x^{2}_{I}
\psi) \right. \nonumber\\
&&-\partial_{I}\left(\partial_{-}hg^{-J}\Delta_{JI}
+\frac{1}{2}\partial_{I}\psi\right) 
-9\partial_{-}\partial_{+}\psi -4\partial_{+}h\partial_{-}h
+{\cal O} (3) + \cdots \Bigg]
\nonumber\\
& = & 2f(\partial_{-}a_{1234}+\partial_{-}a_{5678}) \nonumber\\
&&+\Bigg[-8fh(\partial_{-}a_{1234}-\partial_{-}a_{5678})
+(\partial_{+}a_{1234}\partial_{-}a_{1234}+
\partial_{+}a_{5678}\partial_{-}a_{5678}) \nonumber\\
&& \hspace{6mm}
-\frac{1}{4!}(\partial_{-}a_{1234}\epsilon_{ijkl}\partial_{i}a_{+jkl}
+\partial_{-}a_{5678}\epsilon_{i'j'k'l'}\partial_{i'}a_{+j'k'l'}) \nonumber\\
&&\left. \hspace{6mm}
-\frac{1}{3!}\left((\partial_{-}a_{+ijk})^{2}
+(\partial_{-}a_{+i'j'k'})^{2}\right)
\right] + {\cal O} (3) + \cdots ~ , 
\label{rmp}
\end{eqnarray}
to determine $g^{--}$.  
Inserting the metric component (\ref{gj-}) into (\ref{rmp}),
we obtain an expression of $g^{--}$ up to linear order
\begin{equation}\label{gmmu}
g^{--} =  f^{2}x^{2}_{I}-\frac{1}{\partial^{2}_{-}}\left[
(\partial_{i}^{2}-\partial_{i'}^{2})h-4f\partial_{-}
(a_{1234}+a_{5678})\right] + {\cal O} (2) + \cdots ~ . 
\end{equation}
The expression (\ref{gmmu}) for $g^{--}$ apparently involves the 
four-form gauge field at the linear order, but using (\ref{sdc1})
transmutes the term to the quadratic order, making it irrelevant
when we are allowed to neglect the quadratic and higher 
order terms in $g^{--}$.  In fact, when investigating
the self-dual gauge field equations, the expressions listed in
(\ref{psii}), (\ref{gj-}), and (\ref{gmmu}) are enough
to determine the Lagrangian up to cubic terms.   


For the gauge sector, the self-duality condition (\ref{sdc}) 
reproduces the wave equations (\ref{maxeq}) and, consequently,
we will concentrate on the self-duality condition.
We first solve $F_{-ijkl}$ component equation (\ref{sdc1})
up to the quadratic terms to obtain 
\begin{eqnarray}
a_{1234}& = &a+ 2 \frac{1}{\partial_{-}}(h\partial_{-}a)
+ {\cal O} (3) + \cdots ~ ,
\label{sde1}\\
a_{5678}& = &-a+2 \frac{1}{\partial_{-}}(h\partial_{-}a)
+ {\cal O} (3) + \cdots ~ .
\label{sde2}
\end{eqnarray}
The equation (\ref{sdc1}) actually requires the sum of 
the coefficients of the
second terms in (\ref{sde1}) and (\ref{sde2}) to be four.  
It will turn out
later that this symmetric distribution yields us the 
Lagrangian
that is easiest to quantize; only in this case, the interaction
dependent quadratic contribution to the canonical momentum
is absent.  It is worthwhile to note that the
solutions (\ref{sde1}) and (\ref{sde2}) are analogous to the
nonlocal field redefinition 
\begin{equation}
 \chi \rightarrow  \chi + 2 \frac{1}{\partial_{-}}
( \phi \partial_{-} \chi) \label{yyy}
\end{equation}
introduced in \cite{sug} to investigate the dilaton $\phi$
and axion $\chi$ system. 

The self-duality for the $F_{+-ijk}$ components 
corresponds to the 
Gauss' law constraint part of (\ref{maxeq}).  Utilizing it,
we can express the unphysical 
$(+ijk)$- and  $(+i'j'k')$-components of 
the four-form field in terms of $h$ and $a$:   
\begin{eqnarray}
a_{+ijk}&=&\epsilon_{ijkl}\frac{1}{\partial_{-}}\left[
\partial_{l}a+\left(3h\partial_{l}a
+\left(\frac{\partial_{l}}{\partial_{-}}h
\right)\partial_{-}a-2\frac{\partial_{l}}
{\partial_{-}}(h\partial_{-}a)\right)
\right] + {\cal O} (3) + \cdots ~ ,
\label{aux1}\\
a_{+i'j'k'}&=&\epsilon_{i'j'k'l'}\frac{1}{\partial_{-}}\left[
-\partial_{l'}a+\left(3h\partial_{l'}a
+\left(\frac{\partial_{l'}}{\partial_{-}}h
\right)\partial_{-}a-2\frac{\partial_{l'}}
{\partial_{-}}(h\partial_{-}a)\right) \right] \nonumber \\
 & & + {\cal O} (3) + \cdots ~ ,
\label{aux2}
\end{eqnarray}
where $\epsilon_{1234}=1$ and $\epsilon_{5678}=1$. 
This completes the determination of the relevant 
unphysical fields in terms of the physical fields.


Our next task is the construction of the Lagrangian that
correctly reproduces the equations of motion for $a$ and
$h$ fields. The $(+1234)$-component, 
or equivalently $(+5678)$-component, does the job. 
With the help of self-dual solutions in 
Eqs.~(\ref{sde1})--(\ref{sde2}) and
auxiliary fields in Eqs.~(\ref{aux1})--(\ref{aux2}),
the difference of the two self-duality equations
\begin{eqnarray}
\partial_- \left( F_{+1234}-F_{+5678} \right) =
  \partial_- \left[ 
  \sqrt{-g}(F^{-5678}-F^{-1234}) \right] ~ ,
\end{eqnarray}
leads to an equation for $a$:
\begin{eqnarray}\label{aeq}
\left(\Box a -8f\partial_{-}h\right)
+\left\{\left[-2\frac{\partial_{i}^2}{\partial_{-}}(h\partial_{-}a)
-2\partial_{-}\left(h\frac{\partial_{i}^{2}}{\partial_{-}}a\right)
+3\partial_{i}(h\partial_{i}a)
+\partial_{i}\left(\partial_{-}a\frac{\partial_{i}}{\partial_{-}}h\right)
\right.\right.\nonumber\\
\left. \left. 
+\partial_{-}\left(\partial_{i}a
\frac{\partial_{i}}{\partial_{-}}h\right)
-\partial_{-}\left(\partial_{-}a
\frac{\partial_{i}^{2}}{\partial_{-}^{2}}h
\right)\right] -
\left[i\rightarrow i'\right]\right\}
 + {\cal O} (3) + \cdots = 0 ~ ,
\end{eqnarray}
where $\Box$ denotes d'Alembertian in the pp-wave background 
\begin{equation}
\Box =2\partial_{+}\partial_{-}+\partial^{2}_{I}+f^{2}x^{2}_{I}
\partial^{2}_{-}~ .
\end{equation}
We have verified that (\ref{aeq}) is consistent with 
the second order equation (\ref{maxeq}) up to quadratic 
terms.
Note that the linear terms are a linearized wave equation
for $a$ and the quadratic terms represent cubic interaction of the
type $haa$ at the Lagrangian level. Furthermore,
(\ref{aeq}) is invariant under the transformation
$i \rightarrow i', ~ i' \rightarrow i , ~ a \rightarrow -a , ~
 h \rightarrow -h $ under which both the linear terms 
and the quadratic terms change sign.    
The sum of both equations,
\begin{eqnarray}
F_{+1234}+F_{+5678}=\sqrt{-g}(F^{-5678}+F^{-1234})~,
\end{eqnarray}
reproduces the linear part of Eq.~(\ref{aeq}) 
so it provides a consistency check:
\begin{equation}
\left[h+\partial_{-}h\left(\frac{1}{\partial_{-}}\right)\right]
\left(\Box a -8f\partial_{-}h\right) 
+ {\cal O} (3) + \cdots = 0 ~ .
\end{equation}
From (\ref{aeq}), we can read off the Lagrangian.
The quadratic terms are
\begin{eqnarray}
{\cal L}_{aa}&=& \frac{1}{\kappa^{2}}a\Box a ~ , \label{saa}\\
{\cal L}_{ah}&=& \frac{16}{\kappa^{2}}
 f h\partial_{-}a ~ , 
\label{sah}
\end{eqnarray}
where the choice of overall numerical normalization will be 
explained later.
For the cubic interactions, we obtain 
\begin{eqnarray}
{\cal L}_{aah}&=&
-\frac{1}{\kappa^{2}}\left[4\frac{1}{\partial_{-}}
(h\partial_{-}a)\partial_{i}^{2}a
+3h\partial_{i}a\partial_{i}a
+2\partial_{-}a\partial_{i}a\frac{\partial_{i}}{\partial_{-}}h
-\partial_{-}a\partial_{-}a\frac{\partial_{i}^2}{\partial_{-}^2}h
 \right]
\nonumber\\
&&+\left\{i\rightarrow i',~h\rightarrow 
-h ,  ~  a \rightarrow - a \right\} .
\label{shaa}
\end{eqnarray}
What has been left undetermined so far are the terms involving
only $h$'s, ${\cal L}_{hh}$ and ${\cal L}_{hhh}$, which 
require a separate analysis of the Einstein-Hilbert action
or its equations of motion.


For the gravity sector, since we have already obtained 
all the quadratic and cubic terms involving $a$,
it is sufficient to consider pure gravity case and set $a=0$.
In this case, we can directly adopt the procedures of 
Ref.~\cite{gor}.  The only difference is that the background
metric is that of the pp-wave backgrounds, not the flat metric.
We obtain the following result:
\begin{eqnarray}
{\cal L}_{hh}= \frac{1}{\kappa^{2}}h\Box h, \label{shh}
\end{eqnarray}
and that of the cubic interactions 
\begin{eqnarray}
{\cal L}_{hhh}&=&-\frac{1}{\kappa^{2}}\left[4\frac{1}{\partial_{-}}
(h\partial_{-}h)\partial_{i}^{2}h
+3h\partial_{i}h\partial_{i}h 
+2\partial_{-}h\partial_{i}h\frac{\partial_{i}}{\partial_{-}}h
-\partial_{-}h\partial_{-}h\frac{\partial_{i}^2}{\partial_{-}^2}h
 \right]
\nonumber\\
&&+\left\{i\rightarrow i',~h\rightarrow -h
 , ~a \rightarrow -a \right\} .
\label{shhh}
\end{eqnarray}
In all, the total Lagrangian is given by a sum of five terms
\begin{eqnarray}\label{totl}
{\cal L}&=&{\cal L}_{hh}+{\cal L}_{aa}+{\cal L}_{ah}
+{\cal L}_{hhh}+{\cal L}_{haa}~.
\end{eqnarray}
Several comments should be in order.  First, the numerical
normalization of the ${\cal L}_{hh}$, which determines
the numerical normalization of the ${\cal L}_{hhh}$ term,
has been chosen to reproduce the known quadratic level
equations of motion in Ref.~\cite{metstey}.
Secondly, the mass scale $f$ appears only in the d'Alembertian
part of the quadratic terms.  In fact, the action
$S_{hh}+S_{hhh}$ is of the same form as that of
Goroff and Schwarz~\cite{gor}, expanded up to cubic terms
using (\ref{gaij}), except for the $f$-dependent term in the
d'Alembertian.  The disappearance of the 
$f$-dependence in the cubic terms is consistent with the
case of dilaton-axion system~\cite{sug}.  Thirdly, the resulting
light-cone Hamiltonian from (\ref{totl}) agrees with the 
(zero mode) string field theory analysis of \cite{str} as well.  


Introducing the chiral primary field $s=h+ia$ and its
complex conjugate $\bar{s}=h-ia$, we  
derive the light-cone Hamiltonian via a Legendre transform~:
\begin{equation}
H(\bar{s},s)=H_{2}+H_{3} ~ .
\end{equation}
Specific form of the quadratic Hamiltonian is
\begin{eqnarray}
H_{2}=\frac{1}{\kappa^{2}} \int dx^{-}d^{8}x_{I}\left[
f^{2}x_{I}^{2}\partial_{-}\bar{s}\partial_{-}s
+\partial_{I}\bar{s}\partial_{I}s
+4if(\bar{s}\partial_{-}s-s\partial_{-}\bar{s})
\right] ~ . \label{hquad}
\end{eqnarray}
The cubic interactions do not involve
light-cone time derivatives (a consequence of the choice
made in (\ref{sde1}) and (\ref{sde2})) and the cubic Hamiltonian is 
obtained by a sign flip of the cubic Lagrangian in 
Eqs.~(\ref{shaa}) and (\ref{shhh})~:
\begin{eqnarray}
H_{3}&=&\frac{2}{\kappa^{2}} \int dx^{-}d^{8}x_{I}\Bigg\{
\left[s\partial_{i}s\partial_{i}\bar{s}
-\frac{1}{2}\frac{\partial_{i}}{\partial_{-}}(s\partial_{-}s)
\partial_{i}\bar{s}
-\frac{1}{2}\frac{\partial_{i}}{\partial_{-}}(s\partial_{-}\bar{s})
\partial_{i}s
\right] \nonumber\\ 
&&\hspace{10mm}-\frac{1}{4}\left[s\partial_{i}s\partial_{i}\bar{s}
-\partial_{i}s\partial_{-}\bar{s}\frac{\partial_{i}}{\partial_{-}}s
-\partial_{-}s\partial_{i}\bar{s}\frac{\partial_{i}}{\partial_{-}}s
+\partial_{-}s\partial_{-}\bar{s}\frac{\partial_{i}^{2}}{\partial_{-}^{2}}s
\right]  
\nonumber\\
&&\hspace{10mm}+ \left( i \rightarrow i' ~ , ~
 s \rightarrow -s ~ , ~ \bar{s} \rightarrow - \bar{s} \right) 
 \Bigg\} +\left\{ c.c.\right\},
\label{cham}
\end{eqnarray}
where $c.c.$ stands for the complex conjugation.  The terms
of the type $sss$ and $\bar{s}\bar{s}\bar{s}$ get canceled
when adding the contributions from ${\cal L}_{hhh}$ and
${\cal L}_{haa}$. 
Note that, in order to work
out dynamical issues and structure of constraints systematically,
we should start from the definition of conjugate momenta, which leads to
primary constraints,
\begin{eqnarray}
\Pi_{\bar{s}}&\equiv& \frac{\partial {\cal L}}{\partial(\partial_{+}\bar{s})}
=-\frac{1}{2\kappa^{2}}\partial_{-}s ~ ,\\
\Pi_{s}&\equiv& \frac{\partial {\cal L}}{\partial(\partial_{+}s)}
=-\frac{1}{2\kappa^{2}}\partial_{-}\bar{s} ~.
\end{eqnarray} 
There is no interaction dependent quadratic
contribution to the canonical momenta, indicating that 
the quantization of 
the Hamiltonian up to the cubic terms is straightforward.
For an easier comparison with dilaton-axion $(\tau,\bar{\tau})$ 
system~\cite{sug}, 
we have divided the cubic Hamiltonian (\ref{cham}) into two parts; the
latter terms in the square bracket are Goroff-Schwarz 
terms \cite{gor,sug} that ensure the Lorentz invariance of
the system when $f=0$.  We recall that in order to obtain the 
cubic terms
for the dilaton-axion system, nonlocal
field redefinitions (Eqs.~(2.9)--(2.10) in Ref.~\cite{sug} or
(\ref{yyy})) were 
performed {\it ad hoc} to make sure the absence of
the terms involving light-cone time derivatives 
in the cubic interaction.  In the current analysis of the chiral
primary system, 
similar expressions (\ref{sde1})--(\ref{sde2})
are a perturbative solution of the self-duality condition.
The chiral primary system clearly has an advantage over the
dilaton-axion system when
it comes to the analysis of the higher order interactions. 

We now come to demonstrate our main claim; the interaction
Hamiltonian in (\ref{cham}) is identical to that of the
dilaton-axion system\footnote{A trivial rearrangement of 
$\tau = \phi + i \chi$ to 
remove the factor $i$ in \cite{sug} is necessary.  This way, the real
component of $\tau$ becomes the NS-NS field just as
the case of $s = h + ia$.  In addition, the value of $\sqrt{\det g}$
for the background metric is two in Ref.~\cite{sug} and one
in our case.  We thus multiply (\ref{hquad}) and
(\ref{cham}) by the factor of two.  After the rescaling
$s \rightarrow \frac{\kappa}{\sqrt{2}} s$ and
$\bar{s} \rightarrow \frac{\kappa}{\sqrt{2}} \bar{s}$, we get
the number one in front of (\ref{hquad}) and the factor
$\sqrt{2} \kappa$ in front of (\ref{cham}), consistent
with Eqs.~(2.15) and (2.16) of Ref.~\cite{sug}.} 
given in \cite{sug}, except for the relative
minus sign between the terms involving $\partial_i$ 
and the terms involving $\partial_{i'}$.  The sign 
difference makes the Hamiltonian (\ref{cham}) $Z_2$-invariant
when we assign negative parity to $s$ and $\bar{s}$.  
The quadratic term $H_2$ (\ref{hquad}) indicates that
the energy of the incoming ground state $|s  \rangle_{p^+ >0}$ 
(and the outgoing ground state $\langle \bar{s} |_{p^+ <0}$)  
remains zero when we turn on
$f$, while its complex conjugate incoming state 
$|\bar{s}  \rangle_{p^+ >0}$ (and the outgoing state 
$\langle s |_{p^+ <0}$)and acquires the energy $8f$.  
It is  also instructive
to write down the interaction Hamiltonian in the momentum
basis.  We perform the replacement 
\begin{equation}
\frac{i}{\partial_{-}^{r}}\rightarrow 
 \frac{1}{p^+_r} = \frac{1}{\alpha_{r}}
\end{equation}
for the $r$-th particle to obtain 
\begin{equation}\label{Ham3}
H_3 = \int d\mu_{3}\, h_{3}(\alpha_{r}) s_{1} s_{2} \bar{s}_3 
+ {\rm symmetrization} + (c.c.) ~ .
\end{equation}
Here, the integration measure $d\bar{\mu}_{3}$ is given by 
\begin{equation}\label{measure}
d\mu_{3} = \left( \prod_{r=1}^3 \frac{d\alpha_{r}}{2 \pi}
\frac{ d^{8}p_{r}}{(2\pi)^{8}} \right)
\delta(\Sigma \alpha_{r}) \delta^{8}( \Sigma p_{r})  ~ ,
\end{equation}
and 
\begin{equation}\label{ham3}
h_{3}= \frac{1}{12\kappa^{2}}
\frac{\alpha^4_{1} + \alpha^4_{2} + \alpha^4_{3}}
{\alpha^{2}_{1}\alpha^{2}_{2}\alpha^2_{3}} \left(
  P_{\perp}^2 - P_{\parallel}^2 \right) ~ ,
\end{equation}
where the `momentum' $P$ is expressed as
\begin{equation}
P = \alpha_1 p_2 - \alpha_2 p_1 
= \alpha_2 p_3 - \alpha_3 p_2 = \alpha_3 p_1 
- \alpha_1 p_3 
\end{equation}
in terms of the transverse momentum $p_{r}$ of the $r$-th
particle.  When $\alpha_1 > 0$, $\alpha_2 > 0$ and
$\alpha_3 < 0$ such that $|\alpha_3 | 
= |\alpha_1| + |\alpha_2|$, describing the 
interactions among two incoming chiral
primary states and one outgoing chiral primary state, 
we have 
\begin{equation}
H_{3} \propto
\int d\bar{\mu}_{3}\left[
\frac{(|\alpha_{1}|+|\alpha_{2}|+|\alpha_{3}|)^{4}}{
\alpha_{1}^{2}\alpha_{2}^{2}\alpha_{3}^{2} }
\left( P^2_{\perp}- P^2_{\parallel} \right) s_{1}s_{2}
\bar{s}_{3}
\right] ~ . \label{inh}
\end{equation}
The directions $\parallel$ and $\perp$ denote
the four parallel directions (coming from AdS$_5$) and
the four perpendicular directions (from $S^5$), respectively.  
Written in 
this way, we can compare our results with the (fermionic) 
prefactor part computed for the chiral primary fields
in Refs.~\cite{sug,sug1}.  They 
agree with the light-cone string field theory analysis 
of \cite{str}
and are consistent with string bit theory\footnote{
In the string bit theory framework \cite{bit}, the supergravity 
Hamiltonian
$H_2$ (\ref{hquad}) is obtained in the large $\lambda$ limit,
where the string bits move collectively, satisfying
$x_{\gamma (n)} =
x_{n}$.  The detailed analysis of the relationship between
the light-cone string field theory, string bit theory and
the supergravity will be reported elsewhere \cite{park}.} 
analysis of \cite{bit}, where the
choice of $v_{IJ} = \Delta_{IJ}$ (see (\ref{xxx})) 
was made for the
comparison with the light-cone string field theory.    
An important remaining issue is to understand (\ref{cham})
from the dual gauge theory side \cite{mix}.  The 
string bit theory analysis \cite{bit} indicates that a subtle 
basis change is necessary to find the map between
gauge theory states and string theory states at the
interaction level.

Finally one might  wonder why the interaction Hamiltonian 
for the chiral primary system is quite similar to that for 
the axion-dilaton system as obtained in \cite{sug}. 
For the axion-dilaton system, we just have to replace 
the factor $P^2_{\perp}- P^2_{\parallel}$ in (\ref{inh})
by $P^2_{\perp}+ P^2_{\parallel}$. For the heuristic undestanding
of this difference, one might recall that in the light cone analysis
of the type IIB string theory on the plane wave background, there are 
several choices of the fermionic vacua, which are related by the 
change of the labeling of the states\cite{metstey}. One choice 
is $SO(8)$ invariant vacuum, which corresponds to the axion-dilaton 
state. Chiral primary system corresponds to the vacuum which has 
$SO(4)\times SO(4)$ invariance. Now this difference is reflected in 
the factors mentioned above. For the complete understanding, one has 
to study how the string field theory results obtained in \cite{str}, 
which chooses the axion-dilaton vacuum,  change as one chooses 
different fermionic vaccuum. The result obtained in the paper 
suggests that there should be simple relations between the string 
field theory results defined on the different fermionic vacua.

\section*{Acknowledgement}

We are grateful to K. Choi, B. Chen, and J.-S. Park for 
useful discussions and to
Sangmin Lee for the participation at the early stage
of this work. 
The work of Y. Kiem was supported by Korea Science and 
Engineering Foundation
(KOSEF) Grant R02-2002-000-00146-0 and by 
Korea Research Foundation (KRF) Grant 
KRF-2002-070-C00022. The work of Y. Kim was supported
by KRF Grant KRF-2001-015-DP0082. The work of
J. Park was supported by KRF Grant 
KRF-2002-070-C00022 and by POSTECH BSRI research 
fund 1RB0210601.

\renewcommand{\baselinestretch}{1.20}

\end{document}